\documentclass{aip-cp}

\usepackage[numbers]{natbib}
\usepackage{rotating}
\usepackage{graphicx}

\begin{document}

\title{Compact Tetraquark Interpretation of the $XYZ$ States}

\author[aff1]{Angelo Esposito\corref{cor1}}

\affil[aff1]{Departiment of Physics, Columbia University, 538W 120th Street, New York, NY, 10027, USA}
\corresp[cor1]{Corresponding author: aesposito2458@columbia.edu}

\maketitle

\begin{abstract}
The past decade witnessed the observation of several exotic charmonium-like resonances, some of which are manifestly non-$q\bar q$ mesons. We review one of the most popular interpretations of such states: the compact tetraquark in the constituent diquark-antidiquark picture. Moreover, we discuss some unexplored decay channels which are particularly sensitive to the different phenomenological models and could therefore shed some light on the nature of these fascinating states. Some brief comments on the recently observed pentaquarks are also included.
\end{abstract}

\section{INTRODUCTION}
The constituent quark model provides an incredibly successful phenomenological description of the observed meson and baryon spectrum. However, such a model do not exclude \emph{a priori} the existence of states with a number of valence constituents higher than three. In particular, since the first discovery of the $X(3872)$ at Belle~\cite{Choi:2003ue}, several collaborations observed a large number of states that do not fit the usual meson/baryon picture -- see~\cite{Esposito:2014rxa} for a recent review. Moreover, the observation of charged resonances decaying into charmonia~\cite{Aaij:2014jqa,Ablikim:2013mio,Liu:2013dau} is a compelling proof for the presence of particles with four valence quarks. In Table~\ref{tab1} we report a schematic summary of some of these states.

\begin{table}[t]
\centering
\begin{tabular}{ccccc}
\hline
State & Mass [MeV] & Width [MeV] & $J^{PC}$ & Observed decay modes\\
\hline
$X(3872)$ & $3871.68\pm0.17$ & $<1.2$ & $1^{++}$ & $J/\psi\,\pi^+\pi^-,\,J/\psi\,\pi^+\pi^-\pi^0,\,J/\psi\,\gamma,\,\psi(2S)\,\gamma,\,D\,\bar D^*$\\
$Z_c(3900)^+$ & $ 3888.7\pm3.4$ & $35\pm7$ & $1^{+-}$ & ${(D\,\bar D^*)}^+,\, J/\psi\,\pi^+$ \\
$Z_c(4020)^+$ & $ 4023.9\pm2.4$ & $10\pm6$ & $1^{+-}$ & $h_c\,\pi^+,\,{(D^*\bar D^*)}^+$ \\
$Y(4008)$ & $3891\pm42$ & $255\pm42$ & $1^{--}$ & $J/\psi,\pi^+\pi^-$ \\
$Y(4260)$ & $4250\pm9$ & $108\pm12$ & $1^{--}$ & $J/\psi\,\pi\,\pi,\,J/\psi\,f_0(980),\,Z_c(3900)^+\pi^-,\,X(3872)\,\gamma$ \\
$Y(4360)$ & $4353\pm11$ & $78\pm16$ & $1^{--}$ & $\psi(2S)\,\pi^+\pi^-$ \\
$Z(4430)^+$ & $4478\pm17$ & $180\pm31$ & $1^{+-}$ & $\psi(2S)\,\pi^+,\,J/\psi\,\pi^+$ \\
$Y(4630)$ & $4634^{+9}_{-11}$ & $92^{+41}_{-32}$ & $1^{--}$ & $\Lambda_c^+\bar \Lambda_c^-$ \\
$Y(4660)$ & $4665\pm10$ & $53\pm14$ & $1^{--}$ & $\psi(2S)\,\pi^+\pi^-$ \\
\hline
\end{tabular}
\caption{Summary of the properties of some of the exotic states discovered in the charm sector.}\label{tab1}
\end{table}

Despite the intense experimental effort, the ``zoology'' of these resonances is complicated and we still lack of a comprehensive theoretical framework. In particular, the most popular phenomenological models proposed to explain the internal structure of these particles are the compact tetraquark in the constituent diquark-antidiquark picture (or just tetraquark from now on)~\cite{Maiani:2004vq,Maiani:2014aja}, the loosely bound meson molecule~\cite{Braaten:2003he,Close:2003sg,Swanson:2006st,Cleven:2011gp,Cleven:2013sq}, the hadro-charmonium~\cite{Dubynskiy:2008mq} and the gluonic hybrid~\cite{Guo:2008yz}. 

In the present work we will summarize the main concepts behind the tetraquark model and suggest a decay channel that might help distinguish between the tetraquark and the meson molecule.

\section{THE TETRAQUARK}
\emph{\textbf{Spectroscopy}} --- In the compact tetraquark picture the four constituents are considered as tightly bound together in a diquark-antidiquark configuration, $[Qq_1]_{\bar 3_c}[\bar Q \bar q_2]_{3_c}$, where $Q$ and $q_i$ are respectively a heavy and a light quark and $3_c\,(\bar 3_c)$ is the (anti-)fundamental representation of the $SU(3)_c$ color group. The interaction between the constituents is described by a color-spin Hamiltonian~\cite{Maiani:2004vq,Maiani:2014aja}:
\begin{equation} \label{H}
H=\sum_im_i-2\sum_{i\neq j}\kappa_{ij}\,\vec{S}_i\cdot\vec{S}_j\frac{\lambda_i^a}{2}\frac{\lambda_j^a}{2},
\end{equation}
where $m_i$ are the masses of the constituents, $\kappa_{ij}$ are unknown couplings, $\vec{S}_i$ are spin vectors and $\lambda_i$ are the Gell-Mann matrices. The ground state tetraquarks are taken as eigenstates of the Hamiltonian (\ref{H}). In particular, the possible $S$-wave states obtained combining different diquark and antidiquark spins are
\begin{eqnarray}
&X_0=|0,\bar 0\rangle_0;\quad\quad X_0^\prime=|1,\bar 1\rangle_0;&\quad\quad\quad (J^{PC}=0^{++}), \label{eq2} \\
&X_1=\frac{1}{\sqrt{2}}\left(|1,\bar 0\rangle_1 +|0,\bar 1\rangle_1\right);&\quad\quad\quad (J^{PC}=1^{++}), \\
&X_2=|1,\bar 1\rangle_2;&\quad\quad\quad (J^{PC}=2^{++}), \label{eq4}
\end{eqnarray}
for the neutral ones and
\begin{eqnarray}
&Z=\frac{1}{\sqrt{2}}\left(|1,\bar 0\rangle_1-|0,\bar 1\rangle_1\right);&\quad\quad\quad (J^{PG}=1^{++}), \\
&Z^\prime=|1,\bar 1\rangle_1;&\quad\quad\quad (J^{PG}=1^{++}), \label{eq6}
\end{eqnarray}
for the charged ones. We use the notation $|s,\bar s\rangle_S$ where $s$ and $\bar s$ are the spins of the diquark and antidiquark respectively and $S$ is the total spin (also corresponding to the total angular momentum $J$ in the $S$-wave case). 

Since the couplings $\kappa_{ij}$ are unknown, some kind of \emph{ansatz} is required. In the very first attempt~\cite{Maiani:2004vq} -- the so-called type-I tetraquark -- they were assumed to be the same as in ordinary mesons and baryons and therefore they were extracted from the known hadronic spectrum. Moreover, the $X(3872)$ was used as input to fit the diquark mass. This assumption however has been recently revisited with the introduction of a newer and more successful type-II paradigm~\cite{Maiani:2014aja}. In this most recent picture one instead assumes that the dominant color-spin interactions are those within the tightly bound diquarks and therefore neglects all the couplings but $\kappa_{Qq}=\kappa_{\bar Q\bar q}$. For the sake of brevity we will mostly focus on this latter model.

The $X_1$, $Z$ and $Z^\prime$ states are identified respectively with the $X(3872)$, $Z_c(3900)$ and $Z_c^\prime(4020)$ physical resonances~\cite{Faccini:2013lda}. Starting from the color-spin Hamiltonian one can compute the masses of the different states and in particular show that $M(X_1)=M(Z)$ and $M(Z^\prime)-M(Z)=2\kappa_{cq}$. The first relation is in very good agreement with experiment\footnote{The phenomenological Hamiltonian (\ref{H}) clearly does not contain any information about possible hyperfine splitting.} while from the second one can fit the coupling and find $\kappa_{cq}\simeq67$ MeV. Once this is done it is also possible to estimate the expected masses for the still \emph{unobserved} states:
\begin{equation}
M(X_2)\simeq M(X_0^\prime)\simeq4000\textrm{ MeV};\quad\quad\quad M(X_0)\simeq3770\textrm{ MeV}.
\end{equation}

The authors of~\cite{Maiani:2014aja} also managed to include the observed $J^{PC}=1^{--}$ vector states in the picture (see again Table~\ref{tab1}) by allowing for odd orbital angular momenta between the diquark and the antidiquark. The resulting eigenstates are obtained by adding $L=1,3$ to the states in Equation (\ref{eq2}) to (\ref{eq6}):
\begin{eqnarray}
&Y_1=|0,\bar 0;0,1\rangle_1;\quad\quad Y_2=\frac{1}{\sqrt{2}}\left(|1,\bar 0;1,1\rangle_1+|0,\bar 1;1,1\rangle_1\right); \quad\quad Y_3=|1,\bar 1;0,1\rangle_1;\quad\quad Y_4=|1,\bar 1;2,1\rangle_1; \\
&Y_5=|1,\bar 1;2,3\rangle_1,
\end{eqnarray}
where now we extended the previous notation to include the orbital angular momentum, $|s,\bar s;S,L\rangle_J$. Note that since the present model is specifically thought for the $J=1$ resonances, we ignored states with other angular momenta. If one excludes the $L=3$ state, which is expected at much higher energies, then the $Y_{1,2,4}$ states can be identified with the observed $Y(4008)$, $Y(4260)$ and $Y(4630)$ respectively. Moreover, the structures observed by BESIII in the $\chi_{c0}\omega$~\cite{ Ablikim:2014qwy} and $h_c\pi^+\pi^-$~\cite{Chang-Zheng:2014haa} channels around 4220 MeV -- and hence dubbed $Y(4220)$ -- have been recently interepreted as a possible observation of the last $Y_3$ state~\cite{Faccini:2014pma}.

Lastly, if one also allows radial excitations, then it is fairly natural to describe the $Z(4430)$, $Y(4360)$ and $Y(4660)$ in terms of the $n=2$ levels of the $Z_c(3900)$, $Y(4008)$ and $Y(4260)$ resonances. Indeed, the corresponding mass splittings are $Z(4430)-Z_c(3900)\simeq593$ MeV, $Y(4360)-Y(4008)\simeq350$ MeV and $Y(4660)-Y(4260)\simeq400$ MeV and are perfectly compatible with those observed in ordinary charmonia and bottomonia, \emph{e.g.} $\psi(2S)-J/\psi\simeq589$ MeV, $\chi_{bJ}(2P)-\chi_{bJ}(1P)\simeq360$ MeV and $\chi_{cJ}(2P)-\chi_{cJ}(1P)\simeq437$ MeV.

\vspace{1em}
\noindent\emph{\textbf{Issues}} --- Just like the other ideas proposed during the last decade, the tetraquark model has to face some issues as well. In particular, the main drawback is that the predicted spectrum is over-populated. For example, the resonances corresponding to the $X_0^{(\prime)}$ and $X_2$ states in Equation (\ref{eq2}) and (\ref{eq4}) have not been observed so far.  Moreover, even if one assumes a minimal $SU(2)_f$ flavor symmetry, then we should expect to observe, \emph{e.g.} the charged isospin partners of the $X(3872)$. While the full isospin triplet has been observed for the $Z_c(3900)$~\cite{Xiao:2013iha}, this is not the case for the $X(3872)$. This fact remains unexplained. 

In general, if one assumes the model to be correct, then the absence of a large portion of the predicted resonances should be explained in terms of some kind of selection rules. Having very little information about the actual internal structure of the tetraquark, looking for these rules is a hard task. A promising idea that presents a nice pattern for the observed states relies on a Feshbach resonance mechanism~\cite{Papinutto:2013uya,Pilloni:2014mgn}.

Another issue is that the color-spin Hamiltonian in Equation (\ref{H}) does not provide any information about the dynamics of the system and hence it cannot be used to describe, for example, the decay into charmonia. The standard technique to compute the decay widths for the exotic states is to write the most general matrix element compatible with Lorentz and discrete symmetries and to assign to it a phenomenological coupling constant. In absence of any further information the coupling constant is usually estimated by dimensional analysis and assuming a ``natural size'' for it\footnote{By ``natural size'' we mean the assumption that dimensionless couplings are of order 1.} -- see again~\cite{Esposito:2014rxa} for further details. 

A nice dynamical picture has been recently developed to try to fill this gap~\cite{Brodsky:2014xia,Lebed:2015sxa}. In this framework the tetraquark is seen as a diquark-antidiquark pair flying away from each other and interacting via a spinless Cornell potential until they stop at the classical turning point, $r_Z$. The model then predicts the coupling of the compact tetraquark to charmonia to be proportional to the charmonium propability density computed at $r_Z$, \emph{i.e.} $g\propto\left|\psi_{Q\bar Q}(r_Z)\right|^2$. In the following section we will use both the prescriptions presented here to test the degree of model dependence of our calculation.

\section{THE $Z_c^{(\prime)}\to\eta_c\rho$ CHANNEL}
Having quite a large number of models proposed to describe the observed $XYZ$ states, it would be of great use to select a certain number of observables able to discriminate between different possiblities. It has been recently proposed~\cite{Esposito:2014hsa,Esposito:2015jra} to use the $Z_c^{(\prime)}\to\eta_c\rho$ decay to probe the internal nature of the two charged resonances and, in particular, to distinguish between a compact tetraquark structure and a molecular one. In the following sections we report a summary of those results.

\vspace{1em}
\noindent\emph{\textbf{Compact Tetraquark}} --- Heavy quark spin symmetry implies that the total wave function for a diquark can be factorized as the spinor of the heavy quark times a wave function that depends on the positions of the two quarks and on the spin of the light one, \emph{i.e.} $\psi_{[cq]}=\chi_c\otimes\phi_{[cq]}(\vec r_c,\vec r_q,s_q)$. This approximation holds up to correction $\mathcal{O}(\Lambda_{QCD}/m_c)\simeq0.25$ which, from now on, will be taken as our theoretical error on the tetraquark amplitudes. A similar factorization also holds for transition matrix elements for the decay into charmonia, $\mathcal{A}=\langle\,\chi_{c\bar c}\,|\,\chi_c\otimes\chi_{\bar c}\,\rangle\langle\,\phi_{c\bar c}\,|\,\hat T\,|\,\phi_{[cq][\bar c\bar q]}\,\rangle$. The first factor is nothing but a Clebsh-Gordan coefficient enforcing heavy quark spin symmetry while the second one is a matrix element that depends on the actual dynamics of the process and will have to be determined phenomenologically according to the criteria explained in the previous section. In particular, for the decay of interest, the most general Lorentz invariant matrix elements that also behave properly under parity and charge conjugation are:
\begin{eqnarray}
&\langle\,J/\psi(\eta,p)\,\pi(q)\,|\,Z(\lambda,P)\,\rangle=g_{Z\psi \pi}\lambda\cdot\eta;\quad\quad\quad\langle\,\eta_c(p)\,\rho(\epsilon,q)\,|\,Z(\lambda,P)\,\rangle=g_{Z\eta_c\rho}\lambda\cdot\epsilon; \\
&\langle\,h_c(p,\eta)\,\pi(q)\,|\,Z(\lambda,P)\,\rangle=\frac{g_{Zh_c\pi}}{M_Z^2}\epsilon^{\mu\nu\rho\sigma}\lambda_\mu\eta_\nu P_\rho q_\sigma,
\end{eqnarray}
where $q$, $p$ and $P$ are four-momenta, $\lambda$, $\eta$ and $\epsilon$ are polarization vectors and the $g$'s are strong phenomenological couplings.
As already mentioned, the effective couplings can be estimated by assuming a natural size for them or by working in the framework introduced in~\cite{Brodsky:2014xia} (which we will call ``dynamical tetraquark''). In our calculation we employed both these assumptions and also both the type-I and type-II paradigms.

\vspace{1em}
\noindent\emph{\textbf{Loosely Bound Molecule}} --- The second model we used to compute the branching ratios of interest is the loosely bound meson molecule. In this picture the $Z_c^{(\prime)}$ is seen as a $D^{(*)}\bar D^*$ bound state. In general, the interaction between the $XYZ$ resonances, the light and the heavy mesons is described via the so-called Non Relativistic Effective Field Theory (NREFT)~\cite{Cleven:2011gp,Cleven:2013sq}. This is a non-relativistic limit of the Heavy Quark Effective Theory (HQET) together with Chiral Effective Field Theory ($\chi$EFT). The reader should refer to~\cite{Esposito:2014hsa} and references therein for the complete NREFT Lagrangian.

\begin{figure}[t]
\centering
\includegraphics{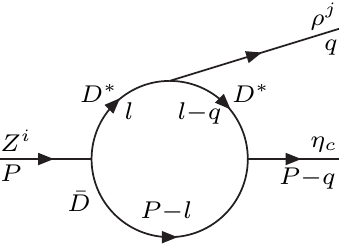} \hspace{3em} \includegraphics{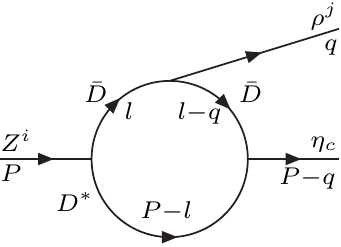} \hspace{3em} \includegraphics{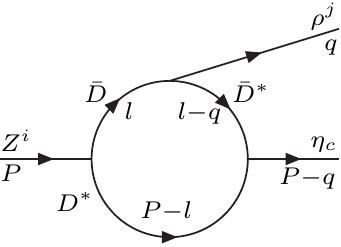} 
\caption{One-loop diagrams computed for the $Z_c$ decay. The charge conjugate diagrams are omitted.} \label{loops}
\end{figure}

\begin{figure}[t]
\centering
\includegraphics{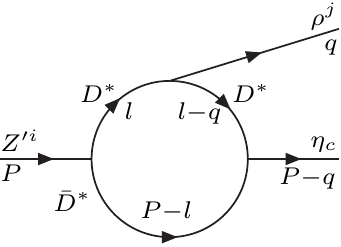} \hspace{3em} \includegraphics{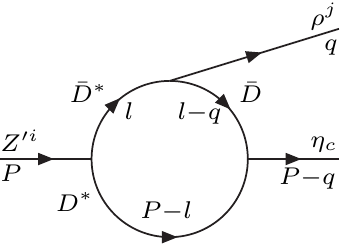}
\caption{Same as in Figure \ref{loops} but for the case of the $Z_c^\prime$.} \label{loopprime}
\end{figure}

The effective theory we just briefly described does not actually contain any information about the molecular structure of the state as it only relies on symmetry arguments. The bound state nature of the $Z_c^{(\prime)}$ is implemented in the model by forcing the resonance to only couple to its own constituents. It then follows that every decay into something different from $D^{(*)}\bar D^*$ must necessarily happen via a heavy meson loop. In Figure~\ref{loops} and \ref{loopprime} we report the one-loop diagrams computed to find the branching fractions in the loosely bound molecule case. 


The theoretical error due to the omission of higher order diagrams can be estimated via a power counting procedure. This technique relies on the fact that, being the molecular states very close to threshold, the typical velocities of the heavy mesons in the loops, $v\simeq\sqrt{|M_Z-2M_D|/M_D}$, are small and hence one can estimate the relevance of the diagrams that have been neglected. In our case, this introduces a 15\% relative error on the single amplitudes.

\begin{figure}[t]
\centering
\includegraphics[width=0.45\textwidth]{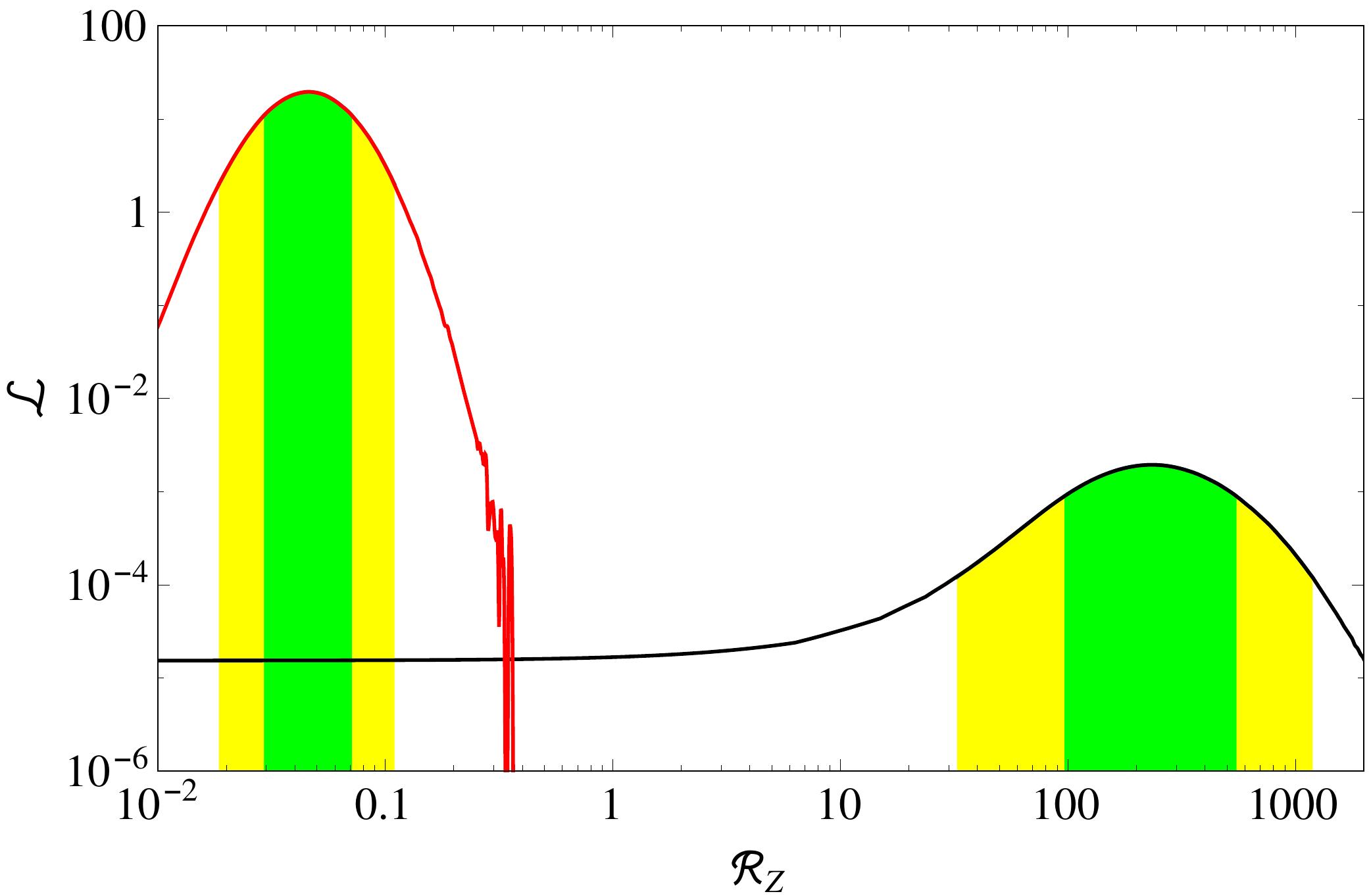} \hspace{2em} \includegraphics[width=0.45\textwidth]{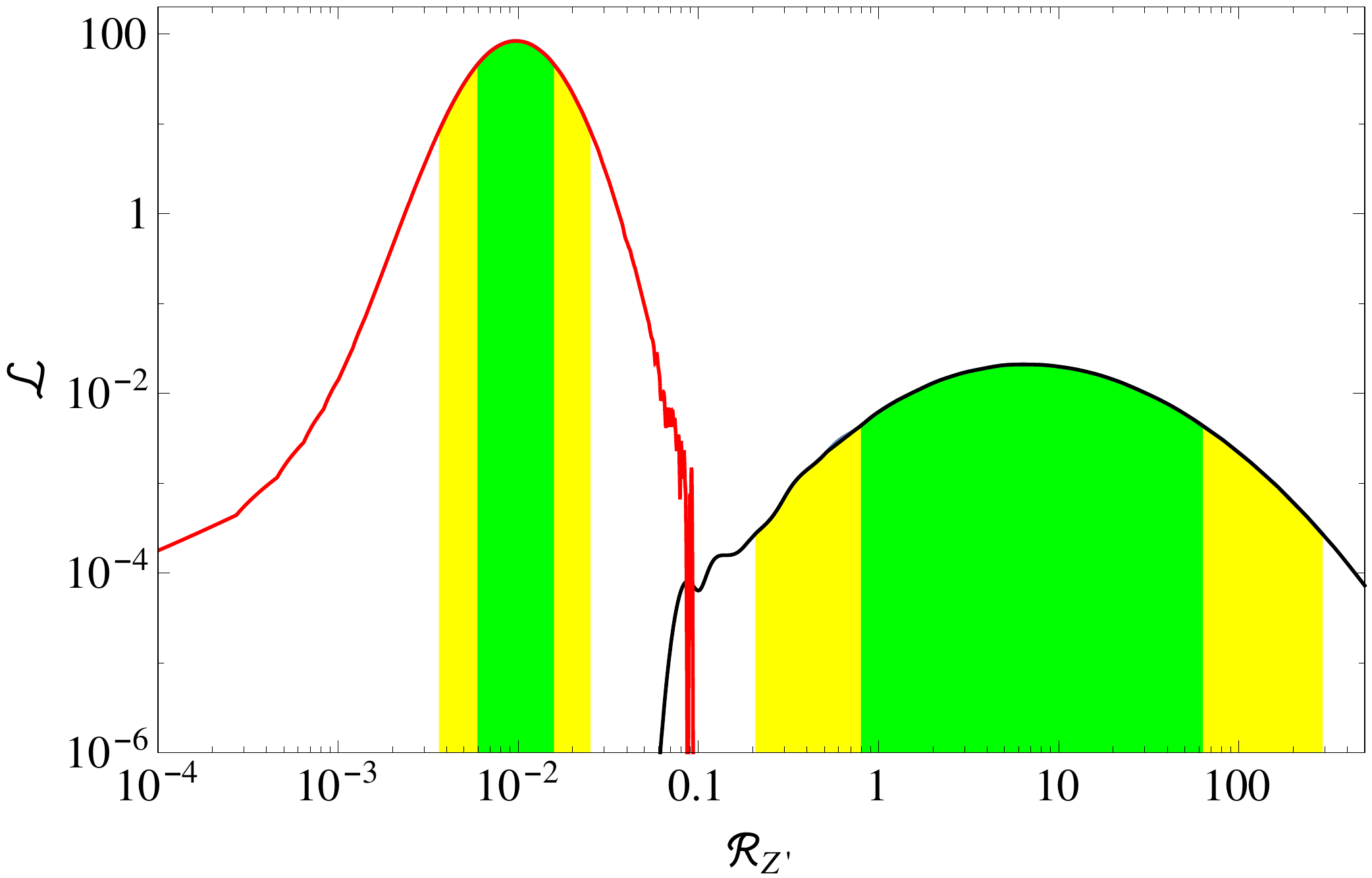}
\caption{Comparison between the likelihood curves for $\mathcal{BR}(Z_c\to\eta_c\,\rho)/\mathcal{BR}(Z_c\to J/\psi\,\pi)$ (left) and $\mathcal{BR}(Z_c^\prime\to\eta_c\,\rho)/\mathcal{BR}(Z_c^\prime\to h_c\,\pi)$ (right). The black and red curves give the predictions according respectively to the dynamical type-I tetraquark and to the meson molecule. The green and yellow band represent the 68\% and 95\% C.L..} \label{comparison}
\end{figure}

\vspace{1em}
\noindent\emph{\textbf{Comparison Between The Models}} --- In Figure~\ref{comparison} we show the comparison between the predictions obtained with the two models. According to the dynamical type-I tetraquark the $Z_c\to\eta_c\,\rho$ decay should be enhanced with respect to the already observed $Z_c \to J/\psi\, \pi$. The opposite is expected in the meson molecule picture and the two predictions are separated by more than $2\sigma$ (95\% C.L.). A similar thing holds for the $Z_c^\prime \to \eta_c\,\rho$ decay with respect to the $Z_c^\prime \to h_c\,\pi$ one. In the latter case, however, the predictions for the type-I and type-II models are the same and hence the result is more model independent. The results obtained assuming no internal dynamics in the tetrquark model are even more separated from the molecular picture. Lastly, in the $Z_c$ case, the type-II paradigm prediction is compatible with the molecular one within $2\sigma$.

\section{BRIEF COMMENTS ON COMPACT PENTAQUARKS}
The LHCb collaboration recently reported~\cite{Aaij:2015tga} the exciting observation of two resonances in the $\Lambda_b$ decay both decaying into a $J/\psi\,p$ final state:
\begin{equation}
\Lambda_b\to\mathbb{P}^+K^-\to(J/\psi\,p)\,K^-.
\end{equation}
In Table~\ref{penta} we report the properties and quantum number of such resonances.
\begin{table}[h]
\centering
\begin{tabular}{cccc}
\hline
State & Mass [MeV] & Width [MeV] & $J^P$ \\
\hline
$\mathbb{P}_{3/2^-}$ & $4380\pm37$ & $205\pm104$ & $3/2^-$ \\
$\mathbb{P}_{5/2^+}$ & $4449.8\pm4.2$ & $39\pm24$ & $5/2^+$ \\
\hline
\end{tabular}
\caption{Properties and quantum numbers of the newly discovered pentaquark resonances.} \label{penta}
\end{table}
One immediately notices that these particles contain a $c\bar c$ pair but also carry a unit of baryon number. In then follows that their minimal quark content must be $c\bar c uud$ and hence they are dubbed \emph{pentaquarks}.

The constituent diquark picture introduced in the previous sections to describe the $XYZ$ mesons~\cite{Maiani:2004vq,Maiani:2014aja} has been used to explain these new states as well~\cite{Maiani:2015vwa}. In this framework, the structure of the two pentaquarks would be $\mathbb{P}_{3/2^-}=\{\bar c[cq_1]_{s=1}[q_2 q_3]_{s=1},L=0\}$ and $\mathbb{P}_{5/2^+}=\{\bar c[cq_1]_{s=1}[q_2 q_3]_{s=0},L=1\}$, where the $q_i$ are light quarks. The negative parity of the first state is then the result of the presence of a single valence antiquark, while the parity ordering instead reflects the opening of the $L=1$ orbital excitation. This pattern is the same as for both ordinary baryons and for the $XYZ$ states and it is a neat signature of the diquark picture.

Lastly, if one considers the mass difference between the $L=0$ and the $L=1$ ordinary baryons it is typically of order $300$ MeV. However, in this case, the lightest $\mathbb{P}_{3/2^-}$ pentaquark also contains a so-called ``bad'' ($s=1$) diquark. It is known from charm and beauty baryon spectra that $s=1$ diquark are around $200$ MeV heavier than the $s=0$ one. Putting everything together, the mass difference of $\sim70$ MeV between the two states is easily explained.

Lastly, the authors of~\cite{Maiani:2015vwa} extended the model to the full $SU(3)_f$ symmetry, also including strangeness.

\section{CONCLUSIONS}
The nature of the so-called $XYZ$ mesons is a now long-standing open problem of hadronic spectroscopy for which a unifying picture is still missing. One of the possible solutions to this puzzle is the compact tetraquark model~\cite{Maiani:2004vq,Maiani:2014aja}, which nicely accommodates the observed exotic states. However, despite the many successes, the model still has to face some severe issues. In particular, the predicted spectrum is more populated than the observed one and hence some hints on possible selection rules are needed~\cite{Papinutto:2013uya,Pilloni:2014mgn}. Moreover, it does not provide any information about the internal dynamics of the system which means that, in order to compute actual decay amplitudes, one often has to rely on crude order-of-magnitue estimates. One possible solution to this last problem has been proposed in~\cite{Brodsky:2014xia,Lebed:2015sxa}.

Another general issue that one has to face on the road towards an explanation of the exotic mesons is the difficulty in determining observables that can neatly distinguish between different ideas. We identified the $Z_c^{(\prime)}\to\eta_c\,\rho$ decay as one of these observables~\cite{Esposito:2014hsa,Esposito:2015jra}. In particular, we showed how in most of the cases the tetraquark picture and the molecular one present predictions which are different by more that $95\%$ C.L.. Particularly enlightening should be the decay of the $Z_c^\prime$ since, in the case of the tetraquark, is type-independent and the predictions are hence much more solid.

Lastly, the discovery of resonances with five-quark content and opposite parity opens another intriguing research direction for non-ordinary mesons and baryons. In particular, the masses and quantum number of these pentaquarks seems to give further strength to the diquark picture~\cite{Maiani:2015vwa}. A complete study of the spectroscopy of these states, however, should wait for the observation of other resonances of similar kind.


\section{ACKNOWLEDGMENTS}
The author would like to thank A.~L.~Guerrieri, F.~Piccinini, A.~Pilloni and A.~D.~Polosa for every useful comment and enlightening discussion.


\nocite{*}
\bibliographystyle{aipnum-cp}%
\bibliography{biblio}%

\end{document}